\documentclass[11pt]{article}
\usepackage{graphicx}

\def\resp#1{} %{ \small (#1) }}

\def\sinphi{\sin(2\beta + \gamma)}

\def\mmiss{m_{\rm miss}}

\def\Dstarp{\Dstar{^+}}
\def\Dstarm{\Dstar{^-}}

\def\btoc{b \rightarrow c \bar u  d}
\def\btou{b \rightarrow u \bar c  d}

\def\btodstpipm{\Bz \rightarrow \Dstarmp\pi^\pm}
\def\btodstpi{\Bz \rightarrow \Dstarp\pi^-}

\def\btodstrhopm{B\rightarrow \Dstarmp\rho^\pm}
\def\btodstrho{\Bz\rightarrow \Dstarp\rho^-}

\def\dt{\Delta t}
\def\dtErr{\sigma_{\dt}}
\def\dz{\Delta z}
\def\dzErr{\sigma_{\dz}}

\def\true{{\rm true}}
\def\dttrue{\dt_\true}

\def\pis{\pi_s}

\def\Dmd{\Delta m_d}

\def\mc{Monte Carlo}

\def\zrec{z_{\rm rec}}
\def\ztag{z_{\rm tag}}

\def\fisher{F}

\def\stag{s_{\rm tag}}
\def\smix{s_{\rm mix}}

\def\dstpi{{\Dstar\pi}}
\def\dstrho{{\Dstar\rho}}
\def\rhopi{{\rho+\pi}}
\def\comb{{\rm comb}}
\def\peak{{\rm peak}}

\def\cont{{q\overline q}}

\def\P{{\cal P}}  % for total PDF
\def\T{{\cal T}}  % for Dt part
\def\M{{\cal M}}  % for mrec part
\def\F{{\cal F}}  % for fisher part
\def\R{{\cal R}}  % for resolution function
\def\A{{\cal A}}  % ARGUS
\def\BG{\hat {\cal G}}% bifurcated Gaussian
\def\G{{\cal G}}  % Resolution function Gaussians

\def\r{r}
\def\Dmt{\Delta m_d t}
\newcommand{\BABARPubYear}    {03}

\newcommand{\BABARConfNumber} {015}
\newcommand{\SLACPubNumber} {10057}
\newcommand{\LANLNumber} {0000}

\input pubboard/babarsym

\long\def\inst#1{\par\nobreak\kern 4pt\nobreak
    {\it #1}\par\vskip 10pt plus 3pt minus 3pt}

\begin{document}
{\pagestyle{empty}

\begin{flushright}
\babar-CONF-\BABARPubYear/\BABARConfNumber \\
SLAC-PUB-\SLACPubNumber \\
hep-ex/\LANLNumber \\
July 2003 \\
\end{flushright}

\par\vskip 5cm

% Title of the paper
\begin{center}
\Large \bf 
Study of Time-Dependent \CP Asymmetries with Partial Reconstruction of 
{\boldmath $\btodstpipm$}
\end{center}
\bigskip

\begin{center}
\large The \babar\ Collaboration\\
\mbox{ }\\
\today
\end{center}
\bigskip \bigskip

% Abstract
\begin{center}
\large \bf Abstract
\end{center}

We present a preliminary measurement of the time-dependent \CP asymmetries in
decays of neutral $B$ mesons to the final states $\Dstarmp\pi^\pm$,
using approximately 
$82$ million $\BB$ events collected by the \babar\ experiment
at the \pep2\ storage ring.  Events containing these decays are
selected with a partial reconstruction technique, in which only the
high momentum $\pi^\pm$ and the low momentum 
pion from the $\Dstarmp$ decay are
reconstructed. The flavor of the other $B$ meson in the event is 
tagged using the information from kaon and lepton candidates.
We measure the time-dependent \CP asymmetry 
${\cal A} = -0.063\pm 0.024~(stat.) \pm 0.017~(syst.)$.
We interpret these results in terms of the angles of the 
unitarity triangle to set a bound on $|\sinphi|$.

\vfill
\begin{center}
Contribution to the 
International Europhysics Conference On High-Energy Physics (HEP 2003),
7/17---7/23/2003, Aachen, Germany
\end{center}

\vspace{1.0cm}
\begin{center}
{\em Stanford Linear Accelerator Center, Stanford University, 
Stanford, CA 94309} \\ \vspace{0.1cm}\hrule\vspace{0.1cm}
Work supported in part by Department of Energy contract DE-AC03-76SF00515.
\end{center}

\newpage
} % end of pagestyle{empty}

% Input author list file
\begin{center}
\small

The \babar\ Collaboration,
\bigskip

%% author list as of 02-Jun-2003 (595 authors)
%
B.~Aubert,
R.~Barate,
D.~Boutigny,
J.-M.~Gaillard,
A.~Hicheur,
Y.~Karyotakis,
J.~P.~Lees,
P.~Robbe,
V.~Tisserand,
A.~Zghiche
\inst{Laboratoire de Physique des Particules, F-74941 Annecy-le-Vieux, France }
% ...

\end{center}\newpage

% The body of the paper starts here

\section{INTRODUCTION}
\label{sec:Introduction}

Measuring the angles of the 
Cabibbo-Kobayashi-Maskawa (CKM)
unitarity triangle~\cite{ref:km} 
will allow us
to overconstrain this triangle and 
to test the Standard Model interpretation of \CP violation
in the quark sector. A crucial step
in this scientific program is the measurement of the
angle $\gamma = \arg{\left(- V_{ud} V_{ub}^\ast/ V_{cd} V_{cb}^\ast\right)}$ 
of the unitarity triangle. 

The neutral $B$ meson decay modes
%~\cite{ref:ft1}
$\Bz \rightarrow {\Dstar}^{\pm} h^{\mp}$, where
$h$ is a light hadron ($\pi, \rho, a_1$), have been
proposed for use in measurements of 
$\sin(2\beta+\gamma)$~\cite{ref:book}, 
where $\beta =  \arg{\left(- V_{cd} V_{cb}^\ast/ V_{td} V_{tb}^\ast\right)}$.
Since the time-dependent \CP asymmetries in these modes are expected to
be of order 2\%, large data samples and multiple decay channels are
required for a statistically significant measurement. 
The technique of partial reconstruction of $D^{*-}$ mesons, in which only 
the soft (low momentum) 
pion $\pi_s$ from the decays $D^{*-} \rightarrow \Dzb \pi_s^- $ 
or $D^{*+}\rightarrow \Dz \pi_s^+$ is reconstructed, 
has already been used to select large samples of $B$ meson 
candidates~\cite{ref:cleo-dstpi}. 

This paper reports the preliminary results of a study 
of \CP-violating asymmetries in 
$\btodstpipm$ decays using the partial reconstruction technique. 
The analysis procedures for the selection of the signal and the 
reconstruction of the decay-time 
difference between the two $B$ mesons in the event 
are essentially the same as those we have 
already applied to the measurement of the
\Bz lifetime~\cite{ref:dstpi-lifetime}.

\section{PRINCIPLE OF THE MEASUREMENT}
\label{sec:Principle}

\begin{figure}[hp]
\begin{minipage}{0.48\textwidth}
\begin{center}
        \includegraphics[width=0.8\textwidth]{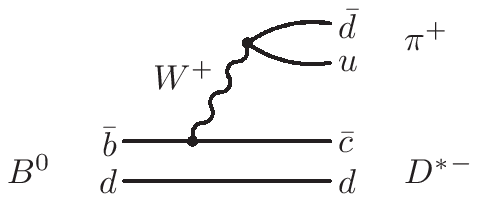}\\
\end{center}
\caption{Feynman diagram for the Cabibbo-favored decay 
$\Bz \rightarrow \Dstarm\pi^+$.}
\label{fig:dstpi-diagrams-1}
\end{minipage}
\hfill
\begin{minipage}{0.48\textwidth}
\begin{center}
        \includegraphics[width=0.8\textwidth]{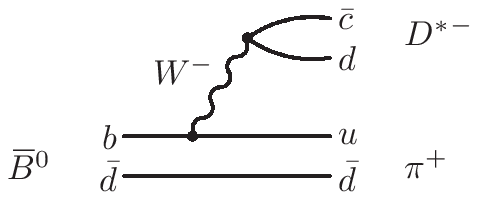}\\
\end{center}
\caption{Feynman diagram for the Cabibbo-suppressed decay 
$\Bzb \rightarrow \Dstarm\pi^+$. 
  }
\label{fig:dstpi-diagrams-2}
\end{minipage}
\end{figure}

The decays $\btodstpipm$ may proceed via a $\btoc$ or a $\btou$
amplitude (see Figs.~\ref{fig:dstpi-diagrams-1}
and~\ref{fig:dstpi-diagrams-2}). Interference between
these amplitudes through $\Bz - \Bzb$ mixing produces time-dependent
\CP-violation observables~\cite{ref:book}. The probability that a state
produced at time~0 as a $\Bz$ or $\Bzb$ decays into the final state
$\Dstarmp\pi^\pm$ at time~$t$ is
%%%
\begin{equation}
\P(\Bz \rightarrow \Dstarmp\pi^\pm)(t) = {1 \over 4 \tau}
        e^{-|t| / \tau} \left[ 1 \pm C \cos(\Dmt)
        + S^\mp \sin(\Dmt)
        \right],
\label{eq:pure-dt-pdf-B}
\end{equation}
%%%
\begin{equation}
\P(\Bzb \rightarrow \Dstarmp\pi^\pm)(t) = {1 \over 4 \tau}
        e^{-|t| / \tau} \left[ 1 \mp C \cos(\Dmt)
        - S^\mp \sin(\Dmt)
        \right],
\label{eq:pure-dt-pdf-BBar}
\end{equation}
%%%%%%%%%%%%%%%%%%%%%%
where $\tau$ is the $\Bz$ lifetime, $\Dmd$  is the $\Bz-\Bzb$
mixing frequency, and we have defined
%%%
\begin{eqnarray}
C &=& {1 - \r^2 \over 1 + \r^2}\, , \nonumber\\
S^\pm &=& {2 \r \over 1 + \r^2}\, \sin(2 \beta + \gamma \pm \delta).
\label{eq:AandB}
\end{eqnarray}
%%%%%%%%%%%%%%%%%%%%%%%
Here 
$\delta$ is an unknown \CP-conserving
phase~\footnote{The definition of $\delta$ is subject to additional $\pi$
terms~\cite{ref:Fleischer} that we ignore, as they are redundant with
the discrete ambiguity $2\beta+\gamma\to 2\beta+\gamma+\pi$, $\delta\to\delta+\pi$.}, 
and 
$r$ is the ratio of the magnitudes of the 
$\btou$ and $\btoc$ amplitudes
\begin{equation}
r = \frac{|A(\Bzb \rightarrow \Dstarm\pi^+)| }
{|A(\Bzb \rightarrow \Dstarp\pi^-) |}.
\end{equation}
Since the $\btou$ amplitude is doubly Cabibbo-suppressed with respect
to the $\btoc$ amplitude, one expects $r \approx 2\%$. 
Due to the small value of $r$, we
use the approximations
%%%
\begin{eqnarray}
C &\approx& 1  , \nonumber\\
S^\pm &\approx& 2 \r \sin(2 \beta + \gamma \pm \delta).
\label{eq:AandB-approx}
\end{eqnarray}
%%%%%%%%%%%
%
In principle, $r$ can be measured from the first two terms 
in the square brackets of Eqs.~\ref{eq:pure-dt-pdf-B} 
and~\ref{eq:pure-dt-pdf-BBar}. 
In practice, this requires sensitivity to terms of order $r^2$,
which is not available with the statistics of current data sets.

%%%%%%%%%%%%%%%%%%%%%%%

Due to the small value of $r$, a large number of signal events is
required in order to observe and measure the small time-dependent \CP
asymmetry.  In the partial reconstruction method, 
the decay $\btodstpipm$ is identified
by reconstructing only the hard (high momentum) 
$\pi^\pm$ and the soft pion $\pis^\mp$
from the decay of the $\Dstarmp$. The four-momentum
of the unreconstructed neutral $D$ meson 
produced in the ${\Dstar}^\mp$ decay
is calculated from the two observed 
tracks and the kinematic constraints relevant for signal decays. 
Partial reconstruction provides a way to obtain very large
single-mode signal samples, by making use of events that cannot be
fully reconstructed.

%%%%%%%%%%%%%%%%%%%%%%%%%%%%%%%%%%%%%%%%%%%%%%%%%%%%%%%%%%%%%%%%%%%%
\section{THE \babar\ DETECTOR AND DATASET}
\label{sec:babar}
The data used in this analysis were collected with the \babar\
detector at the \pep2\ storage ring. The data sample consists of
76.4~fb$^{-1}$ collected on the $\Upsilon(4{\rm S})$ resonance, and
7.6~fb$^{-1}$ collected at an $\epem$ center-of-mass (CM) energy approximately
40~\mev below the resonance peak. Samples of simulated events
with an equivalent luminosity four times larger than the data
were analyzed through the same analysis chain. 

The \babar\ detector is described in detail elsewhere~\cite{ref:babar}.  We
provide a brief description of the main components and their use in this analysis.
Charged particle trajectories are
measured by a combination of a five-layer silicon vertex tracker (SVT)
and a 40-layer drift chamber (DCH) in a 1.5-T solenoidal magnetic
field.  Tracks with low transverse momentum can be reconstructed in the
SVT alone, thus extending the charged-particle detection down to
transverse momenta of $\sim 50$\mevc. Photons and electrons are
detected in a CsI(Tl) electromagnetic calorimeter (EMC), with photon
energy resolution $\sigma_E / E = 0.023 (E/\gev)^{-1/4} \oplus
0.019$. A ring-imaging Cherenkov detector (DIRC) is used for charged
particle identification, augmented by energy loss information
from the SVT and DCH. The instrumented flux return (IFR) is
equipped with resistive plate chambers to identify muons. 

%%%%%%%%%%%%%%%%%%%%%%%%%%%%%%%%%%%%%%%%%%%%%%%%%%%%%%%%%%%%%%%%%%%%%%%
\section{ANALYSIS METHOD}
\label{sec:Analysis}

%%%%%%%%%%%%%%%%%%%%%%%%%%%%%%%%%%%%%%%%%
\subsection{Partial Reconstruction of \boldmath $\btodstpipm$}
\label{sec:partial}

In the partial reconstruction of a $\btodstpipm$ candidate, only the hard
pion track from the $B$ decay and the 
soft pion track $\pi_s$ from the decays $D^{*-}\rightarrow \Dzb \pi_s^-$ 
or  $D^{*+}\rightarrow \Dz \pi_s^+$
are reconstructed. The cosine of the angle between the momenta
of the $B$ and the hard pion in the CM frame is then computed:
%%%%
\begin{equation}
\cos\theta_{Bh} = 
        {M_{{\Dstar}^-}^2 - M_{\Bz}^2 - M_{\pi}^2 + E_{\rm CM} E_h
                        \over
         2 p_B |\vec p_{h}|
        },
\label{eq:cosTheta}
\end{equation}
%%%%%%%%%%%%
where $M_x$ is the nominal mass of particle~$x$~\cite{ref:pdg2002}, 
$E_h$ and $\vec p_h$
are the measured CM energy and momentum of the hard pion, $E_{\rm CM}$ is
the total CM energy of the beams, and 
$p_B = \sqrt{E_{\rm CM}^2/4 - M_{\Bz}^2}$. 
Events are required to be in the 
physical region  $|\cos\theta_{Bh}|<1$.
Given $\cos\theta_{Bh}$ and the measured momenta of the $\pi_h$ and $\pi_s$,
the $B$ four-momentum can be calculated up to
an unknown azimuthal angle $\phi$ around ${\vec p}_{h}$. For every
value of $\phi$, the expected $D$ four-momentum ${\cal
P}_D(\phi)$ is determined from four-momentum conservation, and the 
corresponding
$\phi$-dependent invariant mass 
$m(\phi) \equiv \sqrt{|{\cal P}_D(\phi)|^2}$
is calculated.
We define the missing mass
$\mmiss \equiv {1 \over 2}\left[m_{\rm max} + m_{\rm min}\right]$,
where $m_{\rm max}$ and $m_{\rm min}$ are the maximum and minimum
values that $m(\phi)$ may obtain. In signal events, $\mmiss$ peaks at
the nominal $\Dz$ mass $M_{\Dz}$, with a spread of about 3~\mevcc.
The distribution
for combinatoric background events is significantly broader, making the
missing mass the primary variable for distinguishing signal from
background, as described below.
We use four-momentum conservation to 
calculate the CM $D$ momentum vector with the arbitrary choice
$\phi=0$, and use this variable as described below.

%%%%%%%%%%%%%%%%%%%%%%%%%%%%%%%%%%%%%%%%%%%%%%%%%%%%%%%%%%%%%%%%%%%%%%
\subsection{Backgrounds}
\label{sec:bgd}

In addition to $\btodstpipm$ events, the above procedure yields a
sample containing 
the following kinds of events: 
$\btodstrhopm$; 
peaking $\BB$ background (other than $\btodstrhopm$), 
  defined as pairs of tracks coming from
  the same $B$ meson, with the soft pion originating from the decay of a
  charged $\Dstar$, including contributions from $B
  \rightarrow D^{**} \pi $ as well as non-resonant $B
  \rightarrow D^{*} \pi \pi $ decays;
combinatoric $B$ background, defined as all remaining $\BB$ background events;
continuum $\epem \rightarrow \qqbar$, 
where $q$
represents a $u$, $d$, $s$, or $c$ quark.

%

%%%%%%%%%%%%%%%%%%%%%%%%%%%%%%%%%%%%%%%%%%%%%%%%%%%%%%%%%%%%%%%%%
\subsection{Event Selection}
\label{sec:cuts}

To suppress the continuum background, we select events in which the
ratio of the 2nd to the 0th Fox-Wolfram moment~\cite{ref:R2}, computed
using all charged particles and EMC clusters not matched to tracks, is
smaller than 0.40.
Hard pion candidates are required to be reconstructed with at least
twelve DCH hits. Kaons and leptons are rejected based on
information from the IFR, DIRC, energy loss in the SVT and DCH, or 
the ratio of the
candidate's EMC energy deposition to its momentum ($E/p)$.
We define the $\Dstar$ helicity angle $\theta_{\Dstar}$ to be the
angle between the flight directions of the $D$ and the $B$ in the
$\Dstar$ rest frame, calculated with $\phi = 0$. 
Taking advantage of the longitudinal polarization in signal events,
we suppress background by requiring $| \cos \theta_{\Dstar} |$ to be
larger than 0.4.
All candidates are required to be in the range $1.81 < \mmiss <
1.88$~\gevcc. When multiple candidates are found in the same event,
only the one with the $\mmiss$ value closest to $M_{\Dz}$ is used.

%%%%%%%%%%%%%%%%%%%%%%%%%%%%%%%%%%%%%%%%%%%%%%%%%%%%%55
\subsection{Fisher Discriminant}
\label{sec:fisher}

To further discriminate against continuum events, 
we combine fifteen event
shape variables into a Fisher discriminant~\cite{ref:fisher} $F$.
Discrimination is provided due to the fact that $\qqbar$ events tend to
be jet-like, whereas $\BB$ events have a more spherical energy
distribution. Rather than applying requirements to the variable $F$,
we maximize efficiency by using it in the fits described below.
The fifteen variables are calculated using two sets of particles.
Set~1 includes all tracks and EMC clusters, excluding the hard and
soft pion candidates; Set~2 is composed of Set~1, excluding all tracks
and clusters with CM momentum within 1.25~rad~of the CM momentum of the
$D$, calculated with $\phi=0$. The variables, all calculated in the CM
frame, are 
1) the scalar sum of the momenta of all Set~1 tracks and EMC 
clusters in nine
$20^0$ angular bins centered about the hard pion direction;
2) the value of the sphericity, computed with Set~1;
3) the angle with respect to the hard pion of the sphericity axis, computed
with Set~2;
4) the direction of the particle of highest energy in Set~2
with respect to the hard pion;
5) the absolute value of the vector sum of the momenta of
all the particles in Set~2 ; 
6) the momentum $\vec p_h$ of the hard pion and its polar angle.
%

%%%%%%%%%%%%%%%%%%%%%%%%%%%%%%%%%%%%%%%%%%%%%%%%%%%%%%%%%%%%%%%%%
\subsection{Decay Time Measurement and Flavor Tagging}
\label{sec:deltat}

We define $\zrec$ to be the decay position along the beam axis of the
partially reconstructed $B$ candidate. To find $\zrec$, we fit the
hard pion track with a beam spot constraint in the plane perpendicular to
the beams, the $(x,y)$ plane. The actual vertical beam spot size is
approximately $5~\mu$m, but the constraint 
is taken to be $30~\mu$m in the fit in
order to account for the $B$ flight distance in the $(x,y)$
plane. The soft pion is not used in the fit, since it undergoes
significant multiple scattering.

The decay position $\ztag$ of the other $B$ in the event (the tag $B$)
along the beam axis is obtained from all other tracks in the event,
excluding all tracks whose CM
momentum is within 1~rad of the $D$  CM momentum. The remaining tracks
are fit with a beamspot constraint in the $(x,y)$ plane. The track
with the largest contribution to the $\chi^2$ of the vertex, if
greater than 6, is removed from the vertex, and the fit is carried out
again, until no track fails this requirement. 

We then calculate the decay distance $\Delta z = \zrec - \ztag$, and
the decay-time difference $\dt = \Delta z / (\gamma\beta c)$. The
machine boost parameter $\gamma\beta$ is calculated from the beam energies,
and its average value over the run period is 0.55.
The vertex fits used to determine $\zrec$ and $\ztag$ also yield the
$\Delta z$ error $\dzErr$ which is used to compute the event-by-event 
$\dt$ error $\dtErr$.

The flavor of the tag $B$ is determined from 
lepton and kaon candidates. 
The lepton CM momentum is required to be greater than
1.1~\gevc in order to suppress ``cascade'' leptons originating in
charm decays. We identify electron candidates using $E/p$, and the
Cherenkov angle and number of photons detected in the DIRC.
Muons are identified by the depth of penetration in the IFR.
Kaons are identified using the ionization measured in the SVT and DCH,
and the Cherenkov angle and number of photons detected in the DIRC.
In either the lepton or kaon tagging category, if several
tagging tracks are present, the track used for tagging is the one with 
the largest value of $\theta_{tag}$, the CM-frame angle between the track 
momentum and
the $D$ momentum.
This is done in order to minimize the impact of tracks originating
from the unreconstructed $D$.
If there are both identified leptons and kaons in the event, the event is 
tagged using the lepton tracks only.

We apply the following criteria in order to obtain good $\dt$ resolution:
the $\chi^2$ probability of the $\zrec$ vertex fit must be greater than 0.001;
at least two tracks must be used for the tag $B$ vertex fit;
$\dtErr$ is required to be less than 2~ps;
and $|\dt|$ is required to be less than 15~ps.
To minimize the impact of tracks coming from the unreconstructed $D$,
only tagging leptons (kaons) satifying $\cos \theta_{tag}<0.75$ 
($\cos \theta_{tag}<0.50$) are retained.

%%%%%%%%%%%%%%%%%%%%%%%%%%%%%%%%%%%%%%%%%%%%%%%%%%%%%%%%%%%%%%%%%%%
\subsection{Probability Density Function}
\label{sec:pdf}

The analysis is carried out with a series of unbinned maximum likelihood 
fits performed independently for the lepton-tagged and
kaon-tagged events.
The probability density function (PDF) is a function of the missing
mass $\mmiss$, the Fisher discriminant $F$, the decay time difference $\dt$,
and its error $\dtErr$.

The PDF for on-resonance data is a sum over the PDFs of
the identified event types,
\def\alignHere{&&\kern-0.7cm}
\begin{equation}
\begin{array}{lllll}
\P = f_{\BB} \, \Bigl\{\alignHere f_\rhopi\, (f_\dstpi \, \P_\dstpi \alignHere 
        + (1 - f_\dstpi) \P_\dstrho)  \\
  \alignHere + (1 - f_\rhopi) \Bigl[\alignHere{\kern-0.4cm}
        f_\comb\, \P_\comb + (1 - f_\comb) \P_\peak
                 \Bigr] \Bigr\} + 
        (1-f_{\BB}) P_\cont,
\label{eq:main-pdf}
\end{array}
\end{equation}
%%%%%%%%%
where 
$\P_i$ is the PDF for events of type $i$, and
$f_j$ are relative fractions of events, each limited
to lie in the range $[0,1]$.
Each of the PDFs $\P_i$ is a product of the form
\begin{equation}
\P_i(\mmiss, F, \dt, \dtErr, \stag, \smix) 
        = \M_i(\mmiss) \, \F_i(F) \, \T'_i(\dt, \dtErr, \stag, \smix),
\label{eq:prod-pdf}
\end{equation}
%%%%%%%%%%%%%%%%%%%%%%%
where the variables $\stag$ and $\smix$ are determined by the flavor of
the tag $B$ and the charge of the hard pion
\begin{eqnarray} 
  \stag & = & \left\{\begin{array}{ll}
                    +1, \quad & \mathrm{tag\ }B = \Bz  \\
                    -1, \quad & \mathrm{tag\ }B = \Bzb
                  \end{array} \right. ,\\
  \smix & = & \left\{\begin{array}{ll}
                    +1, \quad & \mathrm{unmixed}  \\
                    -1, \quad & \mathrm{mixed}
                  \end{array} \right. ,
\end{eqnarray}
%%%%%%%%%%
and an event is labeled ``unmixed'' if the hard pion is a 
$\pi^- (\pi^+)$ and the tag $B$ is tagged as a $\Bz (\Bzb)$ and ``mixed''
otherwise. 
The functions $\M_i$, $\F_i$, and $\T'_i$ are described below.
The parameters of $\P_i$ are different for each event type, except
where indicated otherwise.

%%%%%%%%%%%%%%%%%%%%
%%%%% mmiss PDF %%%%
%%%%%%%%%%%%%%%%%%%%

The $\mmiss$ PDF for each event type $i$ is the sum of a bifurcated
Gaussian plus an ARGUS function:
\begin{equation}
\M_i(\mmiss) = f^{\BG}_i\, \BG_i(\mmiss) + (1-f^{\BG}_i) \A_i(\mmiss), 
\label{eq:mmiss-pdf}
\end{equation}
%%%%%%%%%%%%
where $f^{\BG}_i$ is the bifurcated Gaussian fraction. The functions 
$\BG_i$ and $\A_i$ are
\begin{eqnarray}
\BG_i(x) &\propto& \biggl\{ \matrix{
        \exp\left[-(x - M_i)^2 / 2\sigma_{Li}^2\right], & 
                                        x < M_i \cr
        \exp\left[-(x - M_i)^2 / 2\sigma_{Ri}^2\right], & 
                                        x > M_i 
                            }, 
\label{eq:bifur}
\\[0.2cm]
\A(x) 
        &\propto& x \sqrt{1-\left({x /M^A_i}\right)^2}\;
   \exp\left[\epsilon_i \left(1-\left({x / M^A_i}\right)^2\right)\right] 
        \, \Theta(M^A_i - x),
\label{eq:argus}
\end{eqnarray}
%%%%%%%%%%%%
where $M_i$ is the peak of the bifurcated Gaussian, $\sigma_{Li}$ and
$\sigma_{Ri}$ are its left and right widths, $\epsilon_i$ is
the ARGUS exponent, $M^A_i$ is its end point, and the proportionality
constants are such that each of these functions is normalized to unit
area.

%%%%%%%%%%%%%%%%%%%%
%%%% Fisher PDF %%%%
%%%%%%%%%%%%%%%%%%%%

The Fisher discriminant PDF $\F_i$ for each event type is a bifurcated
Gaussian, as in Eq.~\ref{eq:bifur}. The parameter values of $\F_\dstpi$,
$\F_\dstrho$, $\F_\peak$, and $\F_\comb$ are identical.

%%%%%%%%%%%%%%%%%%%%%%%%%
%%%% Convolution PDF %%%%
%%%%%%%%%%%%%%%%%%%%%%%%%

The $\dt$-dependent part of the PDF for events of type $i$ is a
convolution of the form
\begin{equation}
\T'_i(\dt, \dtErr, \stag, \smix) = 
        \int d\dttrue\; \T_i(\dttrue, \stag, \smix) \,
        \R_i(\dt - \dttrue, \dtErr),
\end{equation}
%%%%%%%%%%%%%%%
where $\T_i$ is the distribution of the true decay-time difference
$\dttrue$ and $\R_i$ is a resolution function that accounts for detector
resolution and effects such as systematic offsets in the measured
positions of vertices.
%
%%%%%%%%%%%%%%%%%%%%%%%%%%%%%
%%%% Resolution function %%%%
%%%%%%%%%%%%%%%%%%%%%%%%%%%%%
%
The resolution function for events of type $i$ is the sum of three Gaussians:
\begin{equation}
\R_i(\dt - \dttrue, \dtErr) = 
        f^n_i\, \G^n_i(t_r, \dtErr) + (1 - f^n_i - f^o_i)\, \G^w_i(t_r, \dtErr) 
        + f^o_i\, \G^o_i(t_r, \dtErr),
\label{eq:res}
\end{equation}
%%%%%%%%%%%%%%%
where $t_r$ is the residual $\dt - \dttrue$, and $\G^n_i$, $\G^w_i$, and
$\G^o_i$ are the ``narrow'', ``wide'', and ``outlier'' Gaussians. The
narrow and wide Gaussians have the form
%%%%
\begin{equation}
\G^j_i(t_r, \dtErr) \equiv 
        {1 \over \sqrt{2\pi} \, s^j_i\, \dtErr}
  \exp\left(-\,{\left(t_r - b^j_i\dtErr\right)^2  
        \over 2 (s^j_i\, \dtErr)^2}\right), 
\label{eq:Gaussians}
\end{equation}
%%%%%%%%%%%%%%%%%%
where the index $j$ takes the values $j=n,w$ for the narrow and wide Gaussians,
and $b^j_i$ and $s^j_i$ are parameters determined by fits, as described 
in Sec.~\ref{sec:proc}.
The outlier Gaussian has the form 
%%%%
\begin{equation}
\G^o_i(t_r, \dtErr) \equiv 
        {1 \over \sqrt{2\pi} \, s^o_i  }
  \exp\left(-\,{\left(t_r - b^o_i \right)^2  
        \over 2 (s^o_i)^2}\right),
\label{eq:GaussiansOutlier}
\end{equation}
%%%%%%%%%%%%%%%%%%
where in all fits the values of $b^o_i$ and $s^o_i$ are fixed to 0~ps and
8~\ps, respectively, and are later varied to evaluate systematic errors.

%%%%%%%%%%%%%%%%%%%%%
%%%% Dt_true PDF %%%%
%%%%%%%%%%%%%%%%%%%%%

The PDF $\T_\dstpi(\dttrue, \stag, \smix)$ for signal events
corresponds to Eqs.~\ref{eq:pure-dt-pdf-B} and~\ref{eq:pure-dt-pdf-BBar} 
with Eq.~\ref{eq:AandB-approx} and additional parameters to account for
imperfect flavor tagging.
We define $\omega_{\Bz}$ ($\omega_{\Bzb}$) to be
the mistag probability of signal events whose tag $B$ was tagged as a
$\Bz$ ($\Bzb$), when the tagging track is a daughter of the tag $B$. 
Then $\omega = (\omega_{\Bz} + \omega_{\Bzb})/2$ 
is the average mistag rate, and $\Delta \omega = \omega_{\Bz} - \omega_{\Bzb}$
is the mistag rate difference. 
We further define $\alpha$ to be the probability that the tagging
lepton or kaon is a daughter of the unreconstructed $D$ 
produced in the $\btodstpipm$ decay,
and $\rho$ to be the
probability that this track results in a mixed flavor tag. 
With these definitions, the signal PDF is written as 
\begin{eqnarray}
\T_\dstpi(\dttrue, \stag, \smix)  & = &  
          \frac{1}{4\,\tau}\,e^{-\frac{|\dttrue|}{\tau}}\times
                \nonumber \\   
        & & \left\{ (1-\alpha) \left[\left(1-\stag\,\Delta\omega 
                     \right) \right. \right. \nonumber\\
        & & + \smix\,(1-2\,\omega) \cos(\Delta m
          \dttrue) \nonumber \\
        & & \left. - \stag\,(1-2\,\omega)\,
        S^\pm
                \,\sin(\Delta m \dttrue) \right] \nonumber \\
        & & \left. + \alpha (1+ \smix (1-2 \rho)) \right\} ,
\label{eq:CP-pdf-alpharho}
\end{eqnarray}
where the value $\pm$ in $S^\pm$ is determined by the sign of the product
${\stag \smix}$.
The last term accounts for the tags due to daughters of 
the unreconstructed $D$. 
%%%%%%%%%%%
The parameters $\tau$, $\Delta\omega$, $\omega$, $\Delta m$,
and $S^\pm$ are determined from a fit to the data, as described
below.

The tag $B$ may undergo a $b\rightarrow u \bar c d$ decay, and the kaon 
produced in the subsequent decay of the charmed meson may be used for tagging.
This introduces additional terms, which are not present in
Eq.~\ref{eq:CP-pdf-alpharho}.
To take this effect into account, we use an alternative 
parameterization~\cite{ref:abc} for the kaon tags.
In this
parameterization~\cite{ref:abc}, 
the coefficient of the $\sin(\Delta m \dttrue)$ term
in Eq.~\ref{eq:CP-pdf-alpharho} changes, to give
\begin{eqnarray}
\T_\dstpi(\dttrue, \stag, \smix) & = &  
          \frac{1}{4\,\tau}\,e^{-\frac{|\dttrue|}{\tau}}\times
                \nonumber \\   
        & & \left\{ (1-\alpha) \left[\left(1-\stag\,\Delta\omega 
                     \right) \right. \right. , \nonumber\\
        & & + \smix\,(1-2\,\omega)\, \cos(\Delta m
          \dttrue) \nonumber \\
        & & \left. 
- \left((1-2\,\omega)\, (\stag a + \smix c) 
        + \stag \smix b (1-\stag \Delta\omega) \right)
\sin(\Delta m \dttrue)
                        \right] \nonumber\\
        & & \left. + \alpha (1+ \smix (1-2 \rho)) \right\},
\label{eq:CP-pdf-alpharhoABC}
\end{eqnarray}
where
\begin{eqnarray} \label{math:a} a&\equiv& 2 r\sin(2\beta+\gamma)\cos\delta, \end{eqnarray}
\begin{eqnarray} \label{math:b} b&\equiv& 2 r'\sin(2\beta+\gamma)\cos\delta', \end{eqnarray} 
\begin{eqnarray} \label{math:c} c&\equiv& 2\cos(2\beta+\gamma)(r\sin\delta - r'\sin\delta'). \end{eqnarray}
Here $r'$ describes the effective ratio between the 
magnitudes of the $b\rightarrow u \overline c d$ and 
$b\rightarrow c \overline u d$ amplitudes in the tag side decays, and 
$\delta'$ is the effective strong phase difference between these amplitudes.
This parameterization neglects terms of order $r^2$ and $r'^2$.  
%

%%%%%%%%%%%
%% D*rho %%
%%%%%%%%%%%

We take the $\dt$ PDF parameters of $\btodstrhopm$ events to be
identical to those of the $\btodstpipm$ events except for the 
\CP-violating parameters $S^\pm$, $a$, $b$, and $c$, 
which are set to 0 and are later varied 
to evaluate systematic uncertainties.

%%%%%%%%%%%%%
%% comb BB %%
%%%%%%%%%%%%%

The $\dttrue$ PDF for the combinatoric and the peaking $\BB$
background have the same functional form as
Eq.~\ref{eq:CP-pdf-alpharho} but with independent values for the
parameters.  The parameterization of the $\dttrue$ PDF for the peaking
$\BB$ background has been determined from the Monte Carlo sample.
%%%%%%%%%%%%%%%
%% Continuum %%
%%%%%%%%%%%%%%%

The $\dttrue$ PDF for the continuum background has the functional form
\begin{equation}
\T_{\cont} = f^\delta_\cont\, \delta(\dttrue) T^\delta_\cont + 
        (1 - f^\delta_\cont)\,
        \frac{1}{2\,\tau_\cont}\,e^{-\frac{|\dttrue|}{\tau_\cont}} 
                        T^\tau_\cont,
\label{eq:pdf-cont}
\end{equation}
%%%%%%%%%%%%%%%%%%
where 
\begin{eqnarray}
T^\delta_\cont &=& 1-\stag\,\Delta\omega_\cont + 
        \smix\,(1-2\,\omega^\delta_\cont), \nonumber\\
T^\tau_\cont &=& 1-\stag\,\Delta\omega_\cont + 
        \smix\,(1-2\,\omega^\tau_\cont).
\label{eq:mistag-cont}
\end{eqnarray}
%%%%%%%%%%%%%%%%%

%%%%%%%%%%%%%%%%%%%%%%%%%%%%%%%%%%%%%%%%%%%%%%%%%%%%%%%%%%%%%%%%%%
\section{ANALYSIS PROCEDURE}
\label{sec:proc}

The analysis takes place in four steps, each involving maximum likelihood
fits, carried out simultaneously on the on- and off-resonance data samples:
\begin{enumerate}

\item 
\label{step:1}
Kinematic-variable fit: The parameters of $\M_i(\mmiss)$ 
 and the value of $f_\dstpi$ and
 in Eq.~\ref{eq:main-pdf} are obtained from the \mc\
simulation, conducted with the branching fractions from 
Ref.~\cite{ref:pdg2002}.
Using these parameter values, we fit the data using the PDF in
Eq.~\ref{eq:main-pdf}, but with Eq.~\ref{eq:prod-pdf} replaced by
\begin{equation}
\P_i(\mmiss, F) 
        = \M_i(\mmiss) \, \F_i(F).
\label{eq:prod-pdf-kin}
\end{equation}
%%%%%%%%%%%%%%%%%%%%%%%
The parameters determined in this fit are $f_{\BB}$, $f_\rhopi$, 
and $f_\comb$ in
Eq.~\ref{eq:main-pdf}, the parameters of $\M_\cont(\mmiss)$, 
and those of $\F_i(F)$ for both continuum and $\BB$ events. 
\item 
\label{step:1.5}
$\alpha$ and $\rho$ fit: The kinematic-variable fit is repeated to determine the number of signal 
events above and below the cut on $\cos \theta_{tag}$ 
(see section~\ref{sec:deltat}). These values 
are then used to compute the values of $\alpha$ and $\rho$ in the 
$\Delta t$ PDF (Eq.~\ref{eq:CP-pdf-alpharho}). 
This is done  using values for the efficiencies 
of the cut on $\cos \theta_{tag}$ determined from the Monte Carlo simulation. 
\item 
\label{step:2}
Sideband fit: We fit events in the $\mmiss$ sideband $1.81 < \mmiss <
1.84$~\gevcc to obtain the parameters of the combinatoric $\BB$ PDF
$\T'_{\comb}(\dttrue, \stag, \smix, \dtErr)$.
The PDF in Eq.~\ref{eq:main-pdf} is used in this fit, with $f_\rhopi
= 0$ and $f_\comb = 1$, to account for the fact that the sideband is
populated only by continuum and combinatoric $\BB$ events. The
value of $f_{\BB}$ and the
parameters of the continuum PDF $\T'_{\cont}(\dttrue, \stag, \smix, \dtErr)$
in the sideband are also floating in this fit.

\item 
\label{step:3}
Signal-region fit: Using the parameter values obtained in
the previous steps, we fit the data in the signal
region $1.845 < \mmiss < 1.880$~\gevcc.
This fit determines all the floating parameters of the signal PDF
$\T'_\dstpi(\dttrue, \stag, \smix, \dtErr)$, 
and the parameters of the continuum PDF $\T'_\cont(\dttrue, \stag,
\smix, \dtErr)$ except for
$b^0_\dstpi$, $s^0_\dstpi$, $b^0_\cont$, and $s^0_\cont$ of
Eq.~\ref{eq:GaussiansOutlier}.
\end{enumerate}
In steps~\ref{step:2} and~\ref{step:3} we also fit for a possible
difference between the $\Bz$ and $\Bzb$ tagging efficiencies,
which may be different for each event type.
In order to minimize the possibility of experimenter bias,
step~\ref{step:3} of the analysis is carried out in a ``blind'' manner,
such that the values of $S_\dstpi^\pm$ are hidden from the analysts
until all the systematic errors have been evaluated.

The validity of the analysis procedure has been verified using the
\mc\ simulation in two ways.
First, the use of identical $\dt$ PDFs for $\btodstrhopm$ and $\btodstpipm$
events (except for the
\CP violating parameters), as well as for the combinatoric $\BB$ background
in the sideband and in the signal region, is
validated by comparing the $\dt$ distributions for these event types
in \mc\ by means of Kolmogorov-Smirnov tests. The results of fits to the 
$\dt$ distributions are also compared. In all cases, good agreement 
is observed. 
Second, the entire analysis procedure is
carried out on a \mc\ sample containing four times the number of
events observed in the data.  The values of the parameters obtained
in these \mc\ fits, most importantly, the \CP parameters, are consistent
with the input parameters to within the statistical uncertainties.
In the case of the fit to the lepton-tagged events, a bias due to 
the assumption that events tagged with direct and cascade leptons are
described by the same resolution function is studied using the
full Monte Carlo simulation and a 
fast Monte Carlo technique. This bias is found to be
$\mp 0.012$ for $S_\dstpi^\pm$, and a corresponding correction 
is applied to the
results presented in this paper.

%%%%%%%%%%%%%%%%%%%%%%%%%%%%%%%%%%%%%%%%%%%%%%%%%%%%%%%%%%%%%%%%%%
\section{SYSTEMATIC STUDIES}
\label{sec:Systematics}

The systematic uncertainties on the \CP violation parameters 
($S^\pm$ for events tagged with a lepton candidate and 
$a$, $b$, and $c$ for events tagged with a kaon candidate)
are summarized in Table~\ref{tab:syst} and
described here:

\begin{itemize}
\item[(1)] The statistical errors obtained in the kinematic-variable fit
are propagated to the signal-region $\dt$ fit. This is done
by varying the parameters determined by the kinematic-variable fit, 
taking into account
their correlated errors, repeating the signal-region $\dt$ fit with the 
new parameters, and taking the resulting variation in the \CP violation 
parameters as
a systematic uncertainty.

\item[(2)] The same procedure is applied for the statistical errors 
of the parameters determined in the
sideband fit.

\item[(3)] The parameters of the outlier Gaussian for the signal $\Delta t$
PDF that are fixed in the signal-region fit are varied:
$s^o$ is varied between 6 and 10 ps, and $b^o$ between $-2$ and $2$~ps.
The parameters $\alpha$ and $\rho$ are varied by the statistical
uncertainties determined in their fit. 

\item[(4)] We vary $f_\dstpi$ according to the uncertainties in the branching 
fractions ${\cal B}(\Bz\to D^{*-} \pi^+)$ and  
${\cal B}(\Bz\to D^{*-} \rho^+)$~\cite{ref:pdg2002}, 
and repeat the signal region $\dt$ fit.

\item[(5)] In the signal-region $\dt$ fit we set the values of 
$S_\dstrho^\pm$ to 0. To obtain the systematic error due to this, we vary
$S_\dstrho^\pm$ between $-0.04$ and $+0.04$. The same is done for 
$S_\peak^\pm$ and $S_\comb^\pm$ and the resulting variations
in the \CP parameters are combined
linearly to obtain a conservative uncertainty.

\item[(6)] The values of the neutral $B$ meson lifetime $\tau$ and 
the mixing frequency $\Delta m$ are left free in the fit. The fit is
repeated setting $\Delta m$ to the published value~\cite{ref:pdg2002}.
It is also repeated with $\tau$ set to the value obtained by fitting
the Monte Carlo sample, which is lower than the value in
Ref.~\cite{ref:pdg2002} due to tracks originating from the
unreconstructed $D$~\cite{ref:dstpi-lifetime}.
The resulting variations in the  \CP parameters are combined
in quadrature to obtain the systematic uncertainty. 

\item[(7)] The parameters of the peaking background $\dt$ PDF,
nominally taken from \mc, are varied:
we set their values to the values of the corresponding 
combinatoric background parameters, and alternatively fit the parameters
$\omega_\peak$, $\Delta m_\peak$, and $\tau_\peak$ from the data. 
The largest variation is used as a  systematic error. 
We also fit signal \mc\ with and without $\BB$ backgrounds,
taking the difference as a systematic error.
An additional error accounts for a possible difference between the 
combinatoric $\BB$ parameters in the signal region and the sideband,
evaluated by fitting Monte Carlo events with the
parameters of $\T'_{\comb}$ determined from $\BB$ Monte Carlo events in the
signal-region or sideband.

\item[(8)] The uncertainty due to the beam spot constraint is evaluated by
performing the fit on a sample where the beam spot $y$ position
has been shifted by $20~\mu$m. 

\item[(9)] The $z$ length scale of the detector is determined
with an uncertainty of 0.4\% from the reconstruction of 
secondary interactions with a beam pipe of known length \cite{ref:zscale}.
The uncertainty in the \CP parameters due to the detector $z$ scale 
is evaluated performing the fit on a sample where the
$z$ scale has been varied by 0.4\%.

\item[(10)] The effect of imperfect detector alignment
on a time dependent CP violation measurement is
estimated in a similar study using fully reconstructed decays 
and is assumed to be the same for this analysis.

\item[(11)] The bias due to cascade leptons is evaluated by varying the
parameters of the corresponding PDF. The fraction of cascade
leptons is varied by 3\% and  the mistag rate by 2\%.

\item[(12)] The statistical error of the fit to the 
signal Monte Carlo sample
is added  to the systematic error to account for
possible bias of the analysis procedure.
\end{itemize}

\begin{table}
\caption{The systematic uncertainties on the \CP violation parameters:
$S^\pm$ for events tagged with a lepton candidate, and 
$a$, $b$ and $c$ for events tagged with a kaon candidate.}
\begin{center}
\begin{tabular}{|l|c|c|c|c|c|} 
\hline
Source & $S^-$ error & $S^+$ error & $a$ error & $b$ error & $c$ error  \\
\hline\hline
(1) Kinematic-variable fit statistics   & 0.0005 & 0.0007 & 0.0009      & 0.0004 & 0.0004 \\
(2) Sideband fit statistics             & 0.0007 & 0.0015 & 0.0004      & 0.0003 & 0.0005\\
\hline
(3) Variation of fixed $\dt$ PDF parameters & $<10^{-4}$& $<10^{-4}$& $<10^{-4}$        & $<10^{-4}$& $<10^{-4}$ \\
(4) Uncertainty in branching fractions  & 0.0028 & 0.0028 & 0.0017      & 0.0002 & 0.0033 \\
(5) Uncertainty in bkg \CP parameters   & 0.0096 & 0.0096 & 0.0128      & 0.0073 & 0.0127 \\
(6) Variation of $\tau$ and $\Delta m$  & 0.0012 & 0.0040 & 0.0052      & 0.0017 & 0.0008\\
(7) Effect of $\BB$ background          & 0.0008   & 0.0068  & 0.0045   & 0.0042 & 0.0054\\
\hline
(8) Beam spot                           & 0.0017 & 0.0012 & 0.0007 & 0.0013 & 0.0006\\
(9) $z$ scale                           & 0.0004 & 0.0004 & 0.0002 & $<10^{-4}$& 0.0003\\
(10) Detector alignment                 & 0.0100 & 0.0100 & 0.0100 & 0.0056& 0.0100\\
(11) Cascade lepton bias                & 0.0052 & 0.0052 & - & - &- \\
(12) MC statistics                      & 0.0128 & 0.0128 & 0.0080 & 0.0040 & 0.0090 \\
\hline
\hline
Total                                   & 0.020 & 0.021 & 0.019 & 0.011 &0.020\\
\hline
\end{tabular}
\end{center}
\label{tab:syst}
\end{table}

%%%%%%%%%%%%%%%%%%%%%

%\clearpage
\section{RESULTS OF THE FITS}

\begin{figure}[hp]
\begin{center}
         \includegraphics[width=0.9\textwidth]{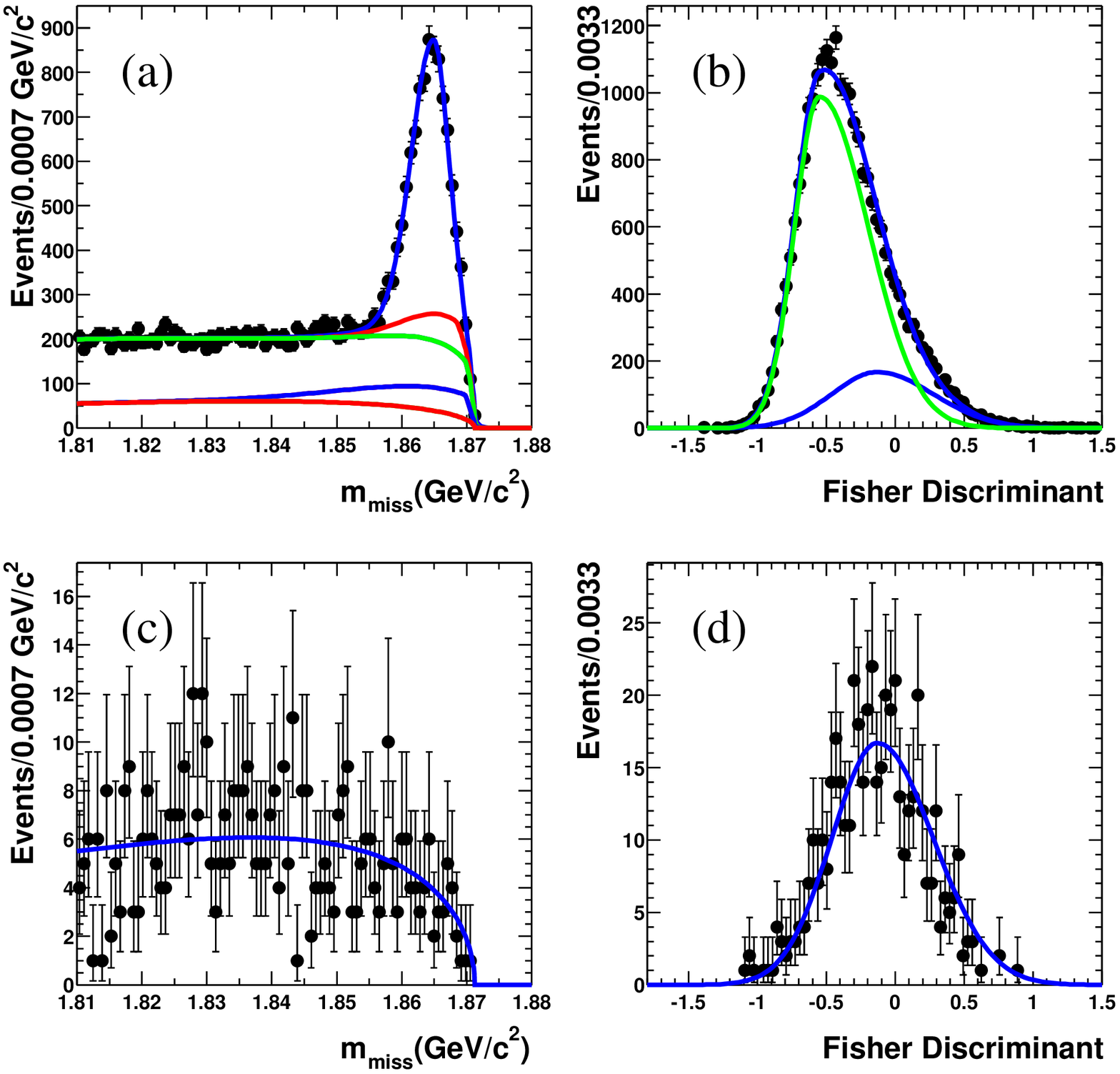}\\
\end{center}
\caption{ Result of the kinematic-variable fit for lepton-tagged
events. The (a)  $\mmiss$ and (b) Fisher
discriminant $\fisher$ distributions for the on-resonance
data are shown by the points with error bars. 
In the $\mmiss$ plot, the overlayed curves show, from bottom
to top, the cumulative contributions of 
continuum, peaking \BB, combinatoric \BB, $\btodstrho$
and $\btodstpi$ events.  In the $\fisher$ plot, the PDFs
for \BB and continuum events are overlaid.  
Plots (c) and (d) show the same distributions for
off-resonance data.  }
\label{fig:data_kin_lepton}
\end{figure}

\begin{figure}[hp]
\begin{center}
         \includegraphics[width=0.9\textwidth]{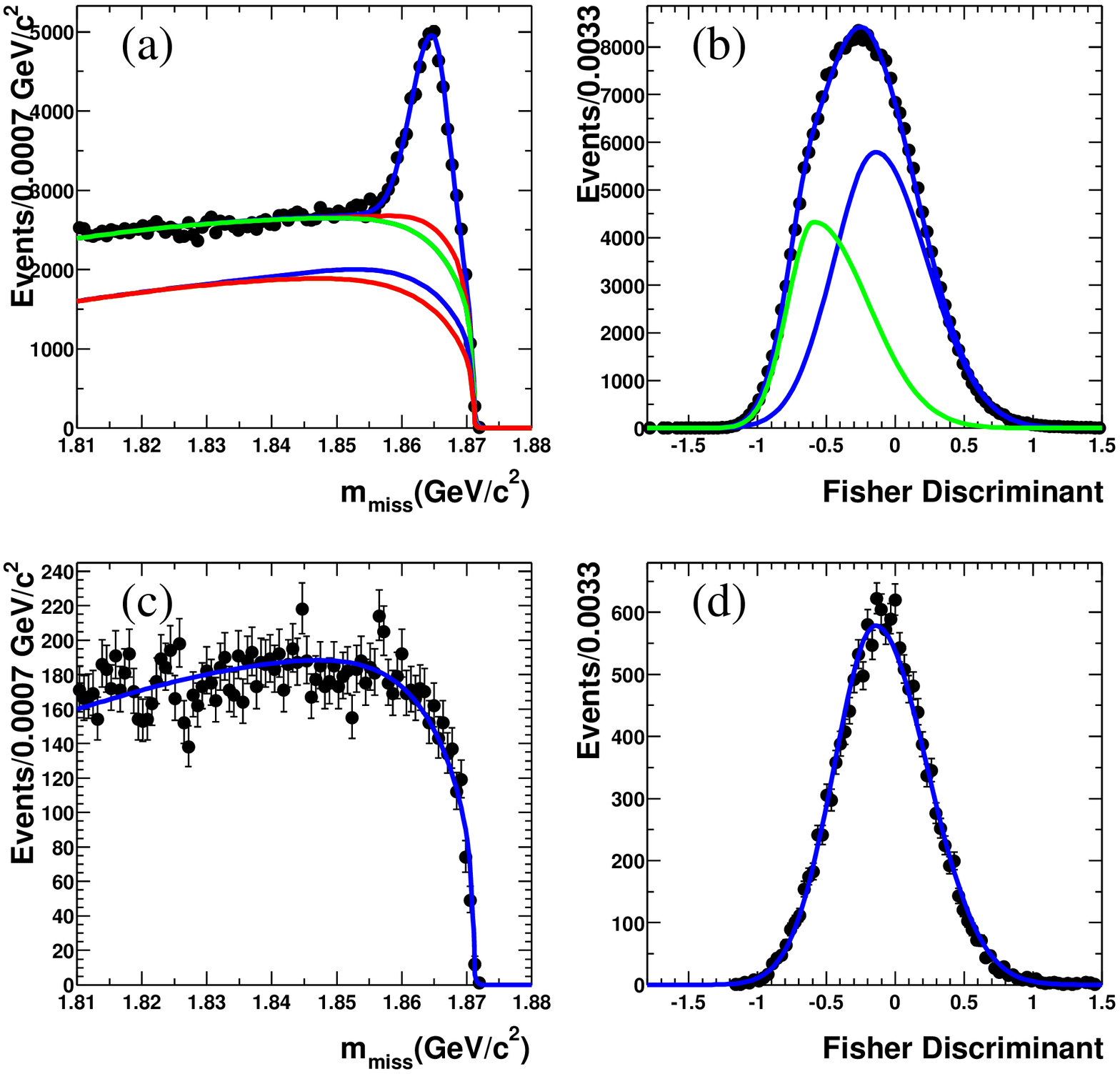}\\
\end{center}
\caption{ Result of the kinematic-variable fit for kaon-tagged
events. 
The (a) $\mmiss$ and (b) Fisher
discriminant $\fisher$ distributions for the on-resonance
data are shown by the points with error bars. 
In the $\mmiss$ plot, the overlayed curves show, from bottom
to top, the cumulative contributions of 
continuum, peaking \BB, combinatoric \BB, $\btodstrho$
and $\btodstpi$ events.  In the $\fisher$ plot, the PDFs
for \BB and continuum events are overlaid.  
Plots (c) and (d) show the same distributions for
off-resonance data.  }
\label{fig:data_kin_kaon}
\end{figure}

The kinematic-variable fit yields $6406 \pm 129$ signal events for the
lepton-tagged sample 
and $ 25157 \pm 323$ signal events for the kaon-tagged sample.
The results of the fit are  shown in Figs.~\ref{fig:data_kin_lepton} 
and~\ref{fig:data_kin_kaon}.
The background in the lepton (kaon) sample is mainly due to $\BB$
(continuum) events.  The kinematic-variable fit is then repeated on
subsamples of the data, separated according to the flavor tag and the
charge of the final state particles (Table~\ref{table:kinfitRes2}).

\begin{table}
\caption{Result of the kinematic-variable fit: number 
of signal events for the entire data sample and for samples separated 
according to
the flavour tag and the charge of the 
final state particles.}
\begin{center}
  \begin{tabular}{|l|c|c|}
    \hline
  & Lepton Tag & Kaon Tag  \\ 
   \hline
All & $6406 \pm 129$ &  $25157 \pm 323 $ \\
$B^0$ tag & $3217 \pm  84 $ & $12821 \pm 232$\\
$\bar B^0$ tag& $3179 \pm  88$ & $12343 \pm 224$\\
$D^{*+} \pi^- $ & $3136 \pm  85$   & $12299 \pm  227 $   \\
$D^{*-} \pi^+ $ & $3269 \pm 84$   &  $12830 \pm 230 $ \\
   \hline
  \end{tabular}
\end{center}
\label{table:kinfitRes2}
\end{table}

\begin{figure}[hp]
\begin{center}
        \includegraphics[width=0.9\textwidth]{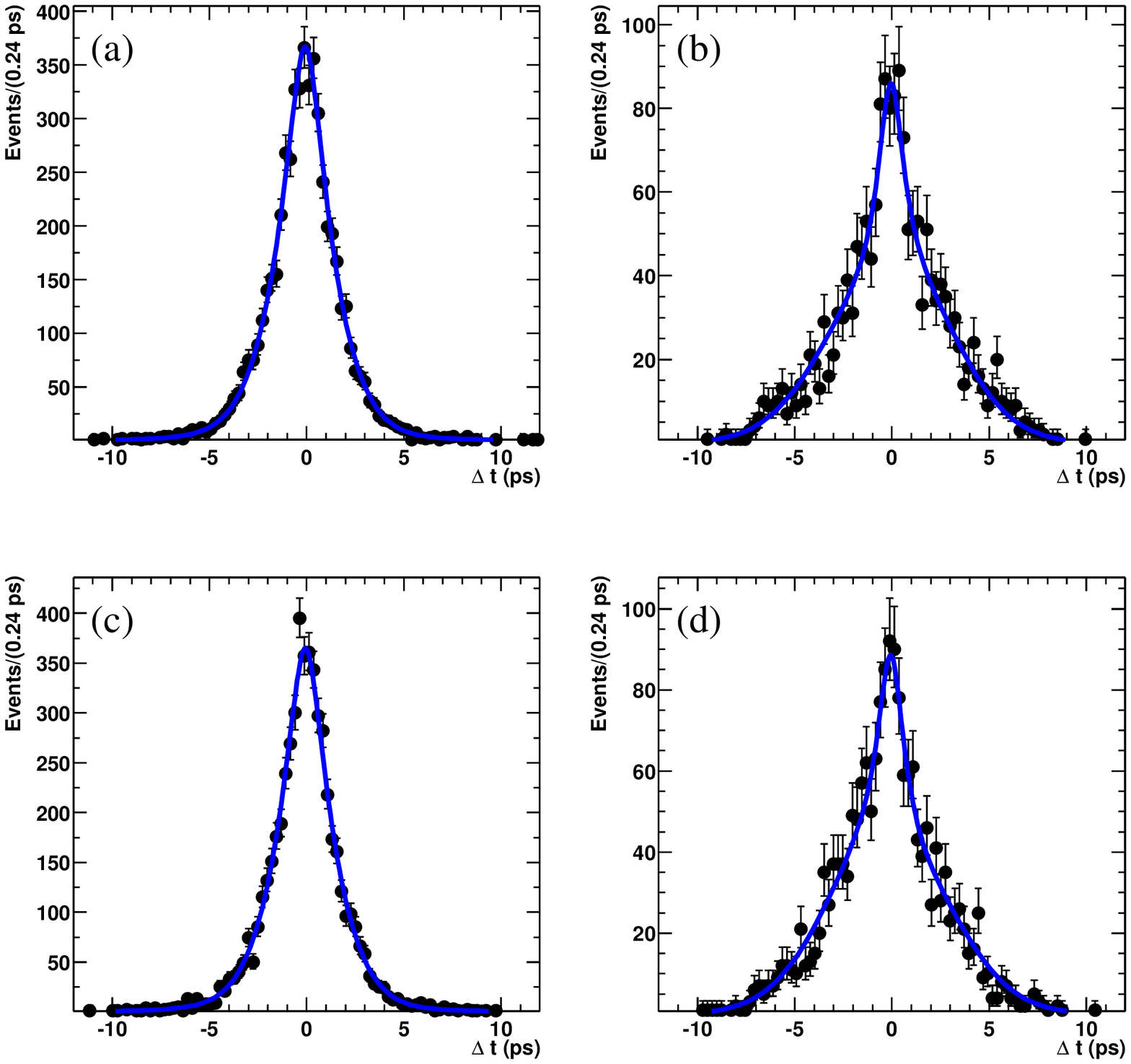}\\
\end{center}
\caption{Result of the $\dt$ fit to the lepton-tagged data in the signal
region. The points with error bars show the $\dt$ distributions for 
(a) $\Bz$-tag unmixed,
(b) $\Bz$-tag mixed,
(c) $\Bzb$-tag unmixed, 
(d) and $\Bzb$-tag mixed
events.
The curves show the PDF with the parameters obtained from
the fit.}
\label{fig:data_sr_lepton}
\end{figure}

\begin{figure}[hp]
\begin{center}
        \includegraphics[width=0.9\textwidth]{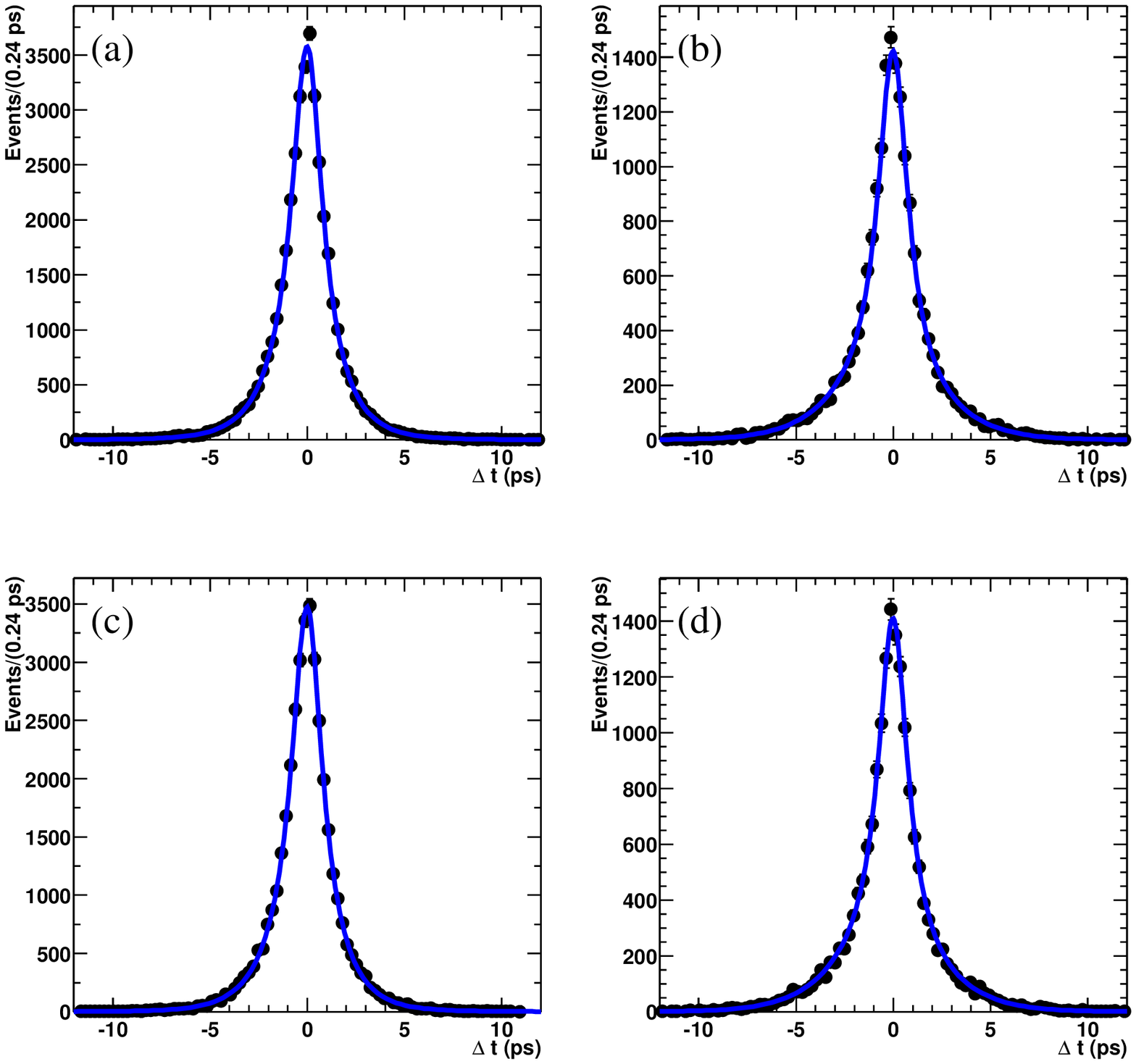}\\
\end{center}
\caption{Result of the $\dt$ fit to the kaon-tagged data in the signal
region. The points with error bars show the $\dt$ distributions for 
(a) $\Bz$-tag unmixed,
(b) $\Bz$-tag mixed,
(c) $\Bzb$-tag unmixed, 
(d) and $\Bzb$-tag mixed
events.
The curves show the PDF with the parameters obtained from
the fit.}
\label{fig:data_sr_kaon}
\end{figure}

The fit to the signal region (Figs.~\ref{fig:data_sr_lepton}
and~\ref{fig:data_sr_kaon}), described in step \ref{step:3} of
Sec.~\ref{sec:proc}, determines $\Delta m$, $\tau$, $\omega$, $\Delta
\omega$, and the \CP parameters of the signal PDF $\T_\dstpi(\dttrue,
\stag, \smix)$. These parameters are $S^\pm$ for lepton tags and $a$, $b$,
$c$ for kaon tags.
Five parameters of the signal resolution function are also determined
by the fit, as are eight continuum parameters: four parameters for the
$\dttrue$ PDF and four parameters for the resolution function.  The
results of the fits on the data are
\begin{eqnarray}
S^+ &=& -0.078 \pm 0.052 ~(stat.) \pm 0.020 ~(syst.), \nonumber\\
S^- &=& -0.070 \pm 0.052 ~(stat.) \pm 0.021 ~(syst.)
\label{eq:results-l}
\end{eqnarray}
for the lepton-tagged events, and
\begin{eqnarray}
a &=& -0.054 \pm 0.032 ~(stat.) \pm 0.019 ~(syst.), \nonumber\\
b &=& -0.009 \pm 0.019 ~(stat.) \pm 0.011 ~(syst.), \nonumber\\
c &=& +0.005 \pm 0.031 ~(stat.) \pm 0.020 ~(syst.)
\label{eq:results-k}
\end{eqnarray}
for the kaon-tagged events.
We note the good agreement between $a$ and
$(S^+ + S^-)/2$.
The largest correlation of $S^+ $ ($S^-$) with any linear combination
of the other parameters floated in the fit is 0.24 (0.26) and the
correlation between $S^+ $  and $S^-$ is $-0.057$.

\begin{figure}[hp]
\begin{minipage}{0.48\textwidth}
\begin{center}
        \includegraphics[width=0.9\textwidth]{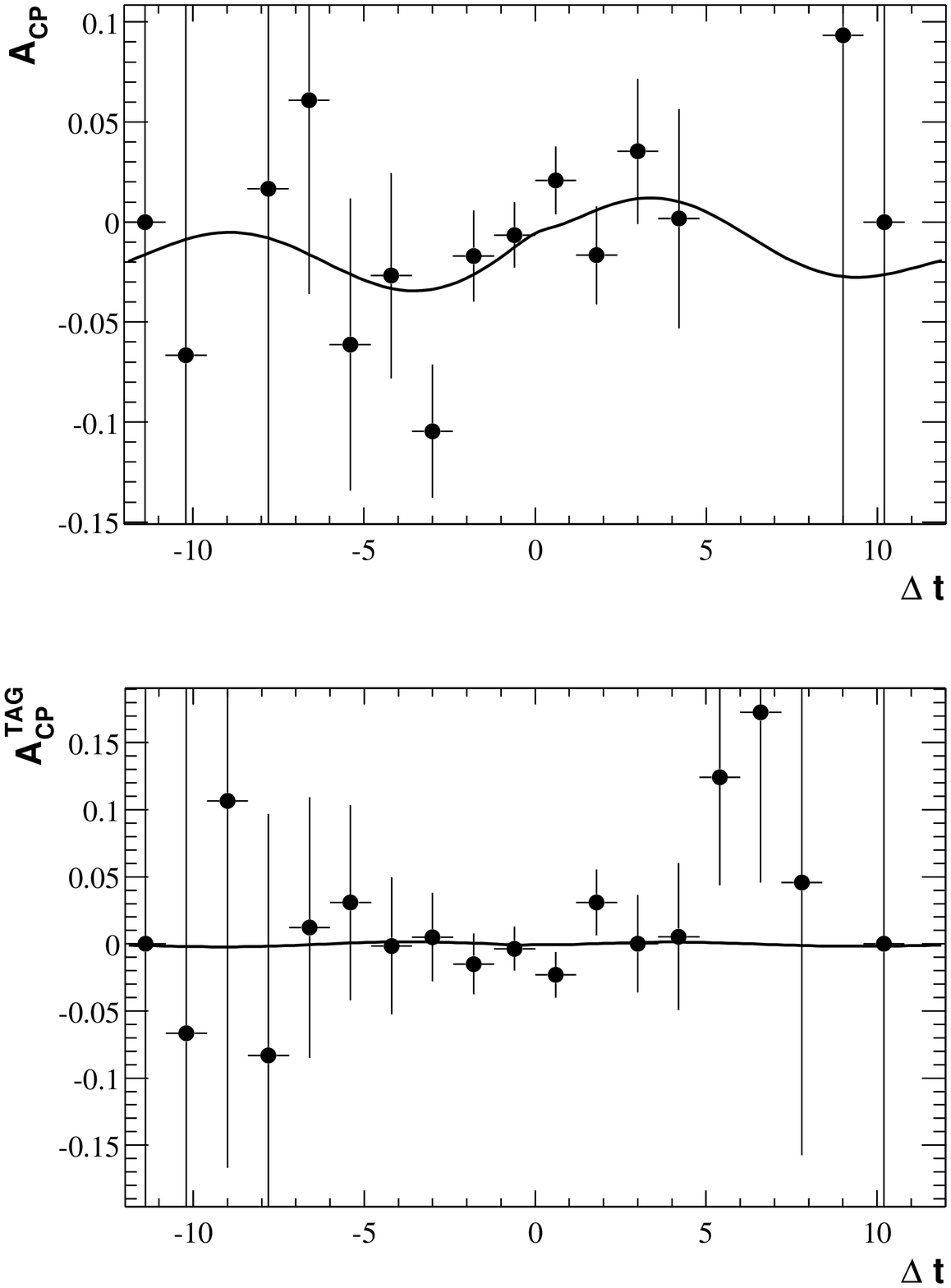}\\
\end{center}
\caption{Raw  asymmetries $\mathcal{A}^{rec}_{\CP}$ and  $\mathcal{A}_{\CP}^{tag}$ (Eq.~\ref{eq:asym}) 
for lepton tags as a function of $\Delta t$. The curves show the 
projection of the fitted PDF.} 
\label{fig:asym_lepton}
\end{minipage}
\hfill
\begin{minipage}{0.48\textwidth}
\begin{center}
        \includegraphics[width=0.9\textwidth]{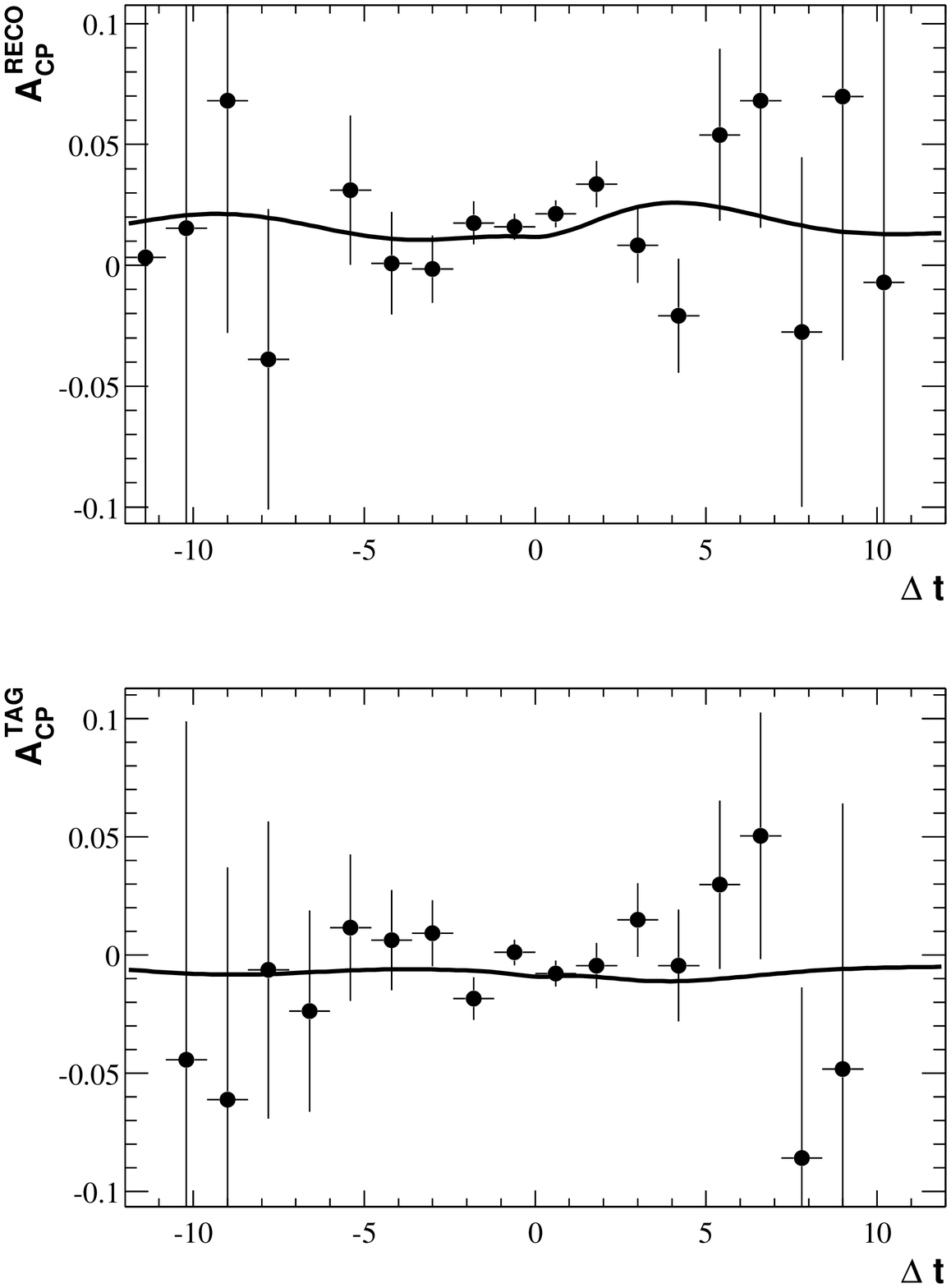}\\
\end{center}
\caption{Raw  asymmetries $\mathcal{A}^{rec}_{\CP}$ and  $\mathcal{A}_{\CP}^{tag}$ (Eq.~\ref{eq:asym}) 
for kaon tags as a function of $\Delta t$. The curves show the 
projection of the fitted PDF. }
\label{fig:asym_kaon}
\end{minipage}
\end{figure}

We define two time-dependent \CP-violating asymmetries from the numbers of 
events observed at time $t$ with specific combinations of
flavor tag and reconstructed final state:
\begin{eqnarray}
\mathcal{A}^{rec}_{\CP} &=& \frac{N({\rm tag~} \Bz, D^{* \pm} 
\pi^{\mp} )(t)
- N( {\rm tag~} \Bzb, D^{*\pm} \pi^{\mp} )(t)}
{N({\rm tag~} \Bz, D^{*\pm} \pi^{\mp} )(t)
+ N({\rm tag~} \Bzb, D^{*\pm} \pi^{\mp} )(t)}, 
        \nonumber\\[2mm]
\mathcal{A}^{tag}_{\CP} &=& \frac{N({\rm tag~} \Bz+ \Bzb, 
D^{*-} \pi^{+} )(t)
- N({\rm tag~} \Bz+\Bzb, D^{*+} \pi^{-} )(t)}
{N({\rm tag~} \Bz+\Bzb, D^{*-} \pi^{+} )(t)
+ N({\rm tag~} \Bz+\Bzb, D^{*+} \pi^{-} )(t)}.
\label{eq:asym}
\end{eqnarray}
In the absence of background and experimental effects,
$ \mathcal{A}^{rec}_{\CP} = - 2 r \sinphi \cos \delta \sin (\Delta m \Delta t )$
and $ \mathcal{A}^{tag}_{\CP} = 2 r' \sinphi \cos \delta' \sin (\Delta m \Delta t )$. 
The asymmetry plots obtained for the data in the signal region are
shown in Fig.~\ref{fig:asym_lepton} for lepton tags and in 
Fig.~\ref{fig:asym_kaon}
for kaon tags. 
As expected, no time-dependent asymmetry is visible 
for $ \mathcal{A}^{tag}_{\CP} $ in the lepton case. 

\section{PHYSICS RESULTS}
\label{section:physics}

   \def\dstpiA{{}}
%    change the above definition to \dstpi to get the subscript back

Combining $a_\dstpiA$ and
$(S^+_\dstpiA + S^-_\dstpiA)/2$ from Eqs.~\ref{eq:results-l} 
and~~\ref{eq:results-k}, accounting for correlated errors, we obtain
\begin{equation}
2 r \sinphi \cos \delta = -0.063 \pm 0.024~(stat.) \pm 0.017~(syst.).
\label{eq:asym-result}
\end{equation}
This measurement deviates from 0 by 2.1 standard deviations,
and is the main result of this analysis.
From the difference $(S^+_\dstpiA -S^-_\dstpiA)/2$,
we obtain
\begin{equation}
2 r \cos(2\beta +\gamma) \sin \delta = 
        -0.004 \pm 0.037~(stat.)\pm 0.020~(syst.).
\end{equation}

We use two different methods for extracting constraints on $|\sinphi|$
from our results. We emphasize that the two methods make use of
different additional information and different assumptions, and are
therefore not directly comparable. These constraints are
interpretations of our experimental results. Each of the methods 
involves defining and minimizing 
a $\chi^2$ function of $\sinphi$ and other parameters. 
The $\chi^2$ functions are 
symmetric under the exchange $\sinphi \rightarrow -\sinphi$.
Due to the large uncertainties and the fact that the minimum value of
the $\chi^2$ may occur at the boundary of the physical region
($|\sinphi|=1$),
the errors naively obtained from the variation of the $\chi^2$ 
functions are not relevant.
In order to give a probabilistic interpretation to the results, we
apply the Feldman-Cousins method~\cite{ref:Feldman} to set limits 
on $|\sinphi|$. 

\begin{figure}[htb]
\begin{center}
        \includegraphics[width=0.7\textwidth]{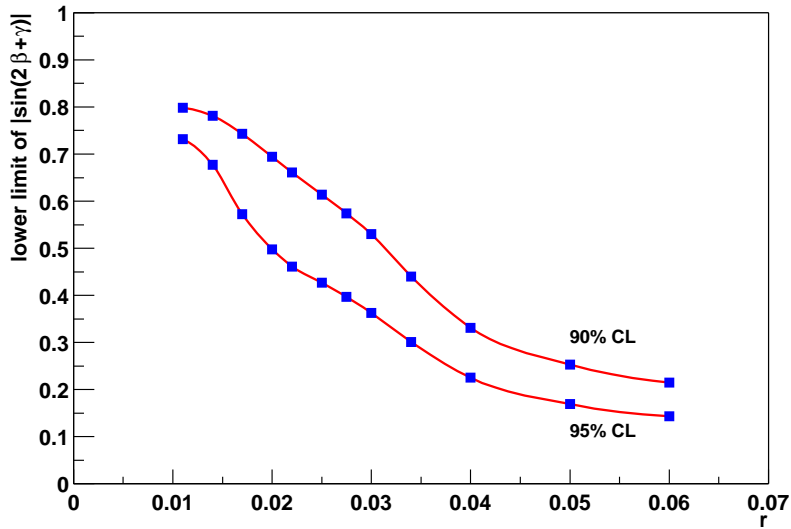}
\end{center}
\vspace*{-0.5cm}
\caption{The 90\% and 95\% CL 
lower limits on $|\sinphi|$ as a function of $r$, 
using no experimental information on $r$.}
\label{fig:limit-vs-r}
\end{figure}

In the first method we make no assumption regarding the value of $r$ and use
no additional experimental information about $r$.
In this method, for different
values of $r$ we minimize the function
\begin{equation}
\chi^2 = \sum_{j,k=1}^3 \Delta x_j V^{-1}_{jk} \Delta x_k ,
\end{equation}
where $\Delta x_j$ is the difference between the result of our
measurement and the theoretical expression for $S^+_\dstpiA$ ($j=1$),
$S^-_\dstpiA$ ($j=2$), and $a_\dstpiA$ ($j=3$), and $V$ is the
measurement error matrix, which is almost diagonal.
The parameters determined by this fit are 
$\sin (2 \beta + \gamma)$, which is limited to lie
in the range $[-1, 1]$, and $\delta$.
The measurements of $b_\dstpiA$ and $c_\dstpiA$ are not used in the fit,
since they depend on the unknown values of $r'$ and $\delta'$.
We generate many parameterized 
MC experiments with the
same sensitivity as reported here for different values of
$\sinphi$ and $r$. The fraction of these experiments in which
$\chi^2(\sinphi) - \chi^2_{min}$ is smaller than in the data is
computed and interpreted as the confidence level
(CL) of the lower limit on $|\sinphi|$. 
The 90\% and 95\% CL limits as a function of $r$ are shown in
Fig.~\ref{fig:limit-vs-r}.
The $\chi^2$ fit determines $|\sinphi|$ up to the twofold ambiguity
$|\sinphi|\leftrightarrow |\cos\delta|$. 
The limits shown in Fig.~\ref{fig:limit-vs-r} are always
the more conservative of the two possibilities.

%%%%%%%%%%%%%%%%%% breco method

\begin{figure}
\begin{center}
        \includegraphics[width=0.6\textwidth]{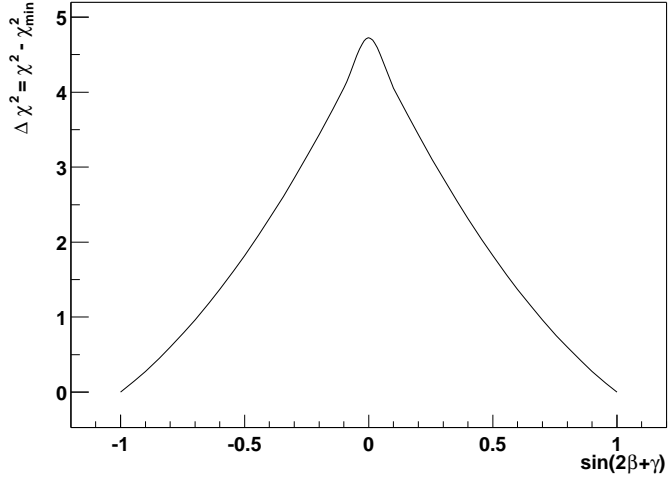}\\
\end{center}
\caption{$\tilde{\chi}^2- \tilde{\chi}^2_{min}$ 
as a function of $\sinphi$, assuming $r = 0.017^{+0.005}_{-0.007}$ 
with an additional 30\% non-Gaussian theoretical uncertainty.}
\label{fig:chi2_sin2bg_res}
\end{figure}

\begin{figure}
\begin{center}
        \includegraphics[width=0.6\textwidth]{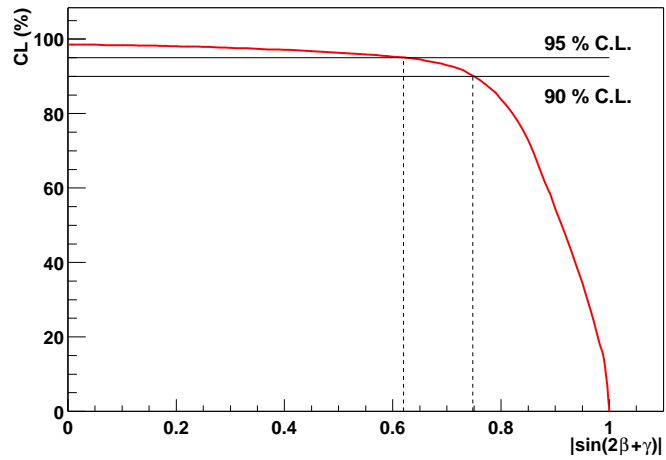}\\
\end{center}
\caption{Confidence level (CL) as a function of the
lower limit on $|\sinphi|$, computed as the fraction
of parameterized MC experiments where 
$\tilde{\chi}^2(\sinphi) - \tilde{\chi}^2_{min}$ is larger 
than in the data, with the assumptions of Fig.~\ref{fig:chi2_sin2bg_res}. 
The vertical lines 
correspond to 90\% and 95\% confidence level and lead 
to the bound $|\sinphi|>0.75~(0.62)$ at the 90\%~(95\%) CL. 
  }
\label{fig:chi2_FC}
\end{figure}

In the second method we assume that $r$ may be estimated from
\begin{equation}
r= \tan\theta_C \, \sqrt{
        {\cal B}(B^0\rightarrow {\Dstar}_s^{+} \pi^-) \over 
        {\cal B}(B^0\rightarrow {\Dstar}^{-} \pi^+)} \, 
        {f_{\Dstar} \over f_{\Dstar_s}}
\end{equation}
where $\theta_C$ is the Cabibbo angle. 
Using the branching fractions 
${\cal B}(B^0\rightarrow {\Dstar}^{-} \pi^+) = 
  (0.276 \pm 0.021)\%$~\cite{ref:pdg2002},
${\cal B}(B^0\rightarrow {\Dstar}_s^{+} \pi^-) = 
  (1.9 {+1.2 \atop -1.3} \pm 0.5) \times 10^{-5}$~\cite{ref:Dspi}
and the ratio of decay constants 
${f_{\Dstar_s} \over f_{\Dstar}} = 1.10 \pm 0.02$~\cite{ref:dec-const}
yields
\begin{equation}
r = 0.017^{+0.005}_{-0.007}.
\end{equation}
An additional
non-Gaussian 30\% relative error is associated with the theoretical assumptions
involved in obtaining this value. To carry out this method, we minimize
\begin{equation}
\tilde{\chi}^2 = \sum_{j,k=1}^3 \Delta x_j V^{-1}_{jk} \Delta x_k 
        + \Delta^2(r),
\label{eq:r}
\end{equation}
where the function
\begin{equation}
\Delta^2(r) = \left\{\matrix{
    \left({r - r_0 \over 0.005}\right)^2 & & (r - r_0) / r_0 > 0.3,  \cr
                                      0  & & |r - r_0| / r_0 \le 0.3, \cr
    \left({r - r_0 \over 0.007}\right)^2 & & (r - r_0) / r_0 < -0.3}
        \right.
\end{equation}
accounts for the 30\% theoretical error and the Gaussian experimental error
around the central value $r=r_0$, Eq.~\ref{eq:r}.
In addition to $\sin (2 \beta + \gamma)$ and $\delta$, the parameter
$r$ is also determined by this fit.
The minimum of $\tilde{\chi}^2$ occurs at $|\sinphi| = 1.0 $,
$\delta= 0$, and $r= 0.024$, and at a value of $\tilde{\chi}^2_{min} = 0.54$
for one degree of freedom.
The value of $\tilde{\chi}^2- \tilde{\chi}^2_{min}$ as a function of $\sinphi$ 
and the resulting
Feldman-Cousins confidence level curve are shown in Figs.~\ref{fig:chi2_sin2bg_res}
and~\ref{fig:chi2_FC}. 
This method yields the limits 
$|\sinphi|> $0.88 at 68\% CL,
$|\sinphi|> $0.75 at 90\% CL, and
$|\sinphi|> $0.62 at 95\% CL.  

\section{SUMMARY}
\label{sec:Summary}

We present preliminary results of a study of time-dependent \CP
asymmetries in the $\btodstpipm$ decay channels using the partial
reconstruction method. The time-dependent \CP asymmetry that we
measure, 
\begin{equation}
2 r \sinphi \cos \delta = -0.063 \pm 0.024~(stat.) \pm 0.017~(syst.),
\label{eq:summary-result}
\end{equation}
is different from 0 by 2.1 standard deviations. This asymmetry
does not depend on assumptions regarding $r$, the ratio of the magnitudes
of the $\btou$ and $\btoc$ amplitudes contributing to this decay.  
We present model-independent bounds on $|\sinphi|$ as a
function of $r$.
With some assumptions regarding $r$, our results can be
interpreted as a limit on the combination of CKM angles $2 \beta +
\gamma$, $|\sinphi|>0.75~(0.62)$ at the 90\%~(95\%) CL.

\section{ACKNOWLEDGMENTS}
\label{sec:Acknowledgments}

% Standard acknowledgments paragraph; must always be included.
We are grateful for the 
extraordinary contributions of our \pep2\ colleagues in
achieving the excellent luminosity and machine conditions
that have made this work possible.
The success of this project also relies critically on the 
expertise and dedication of the computing organizations that 
support \babar.
The collaborating institutions wish to thank 
SLAC for its support and the kind hospitality extended to them. 
This work is supported by the
US Department of Energy
and National Science Foundation, the
Natural Sciences and Engineering Research Council (Canada),
Institute of High Energy Physics (China), the
Commissariat \`a l'Energie Atomique and
Institut National de Physique Nucl\'eaire et de Physique des Particules
(France), the
Bundesministerium f\"ur Bildung und Forschung and
Deutsche Forschungsgemeinschaft
(Germany), the
Istituto Nazionale di Fisica Nucleare (Italy),
the Foundation for Fundamental Research on Matter (The Netherlands),
the Research Council of Norway, the
Ministry of Science and Technology of the Russian Federation, and the
Particle Physics and Astronomy Research Council (United Kingdom). 
Individuals have received support from 
the A. P. Sloan Foundation, 
the Research Corporation,
and the Alexander von Humboldt Foundation.

\end{document}